\documentclass[%
reprint,
 showpacs,
 aps,
prl,
 longbibliography
]{revtex4-1}

\usepackage{amsmath}
\usepackage{amsthm}
\usepackage{dsfont}
\usepackage{mathbbol}
\usepackage{bbold}
\usepackage{mathrsfs}
\usepackage{graphicx}
\usepackage{dcolumn}
\usepackage{bm}
\usepackage{comment}
\usepackage[usenames]{color}
\usepackage[normalem]{ulem}
\usepackage[makeroom]{cancel}

\usepackage[per-mode=symbol]{siunitx}   
\usepackage[version=3]{mhchem}
 
\DeclareSIUnit\intensity{\watt\per\centi\meter\squared}
\DeclareSIUnit\fieldstrength{\volt\per\centi\meter}
\DeclareSIUnit\kfieldstrength{k\volt\per\centi\meter}
\DeclareSIUnit\energy{cm^{-1}}

\newcommand{\ket}[1]{\left|#1\right\rangle}
\newcommand{\ketx}[1]{\left|#1\right\rangle_X}
\newcommand{\bra}[1]{\left\langle #1\right|}

\newcommand{\ie}{i.\,e.}%

\newcommand{\tripletpp}{\ensuremath{{{}^3\text{P}_2}}}
\newcommand{\triplets}{\ensuremath{{{}^3\text{S}}}}
\newcommand{\doublets}{\ensuremath{{{}^2\text{S}}}}
\newcommand{\singlets}{\ensuremath{{{}^1\text{S}}}}

\newcommand{\uoft}{\affiliation{Chemical Physics Theory Group, Department of Chemistry, and Center for Quantum Information and Quantum Control, University of Toronto, Toronto, ON M5S 3H6, Canada}}

\begin{document}

\title{Coherent Control of Penning and Associative Ionization: Insights from Symmetries}

\author{Juan J.\ Omiste}
\author{Johannes Flo{\ss}}
\author{Paul Brumer}\uoft

\date{\today}
\begin{abstract}
Coherent control of reactive atomic and molecular collision processes remains elusive experimentally due to quantum interference-based requirements. Here, with insights from symmetry conditions, a viable method for controlling Penning and Associative ionization in atomic collisions is proposed. Computational applications to He$^*({}^3\text{S})$-Li(${^2\text{S}}$) and Ne$^*{}(^3\text{P}_2$)-Ar($^1\text{S}_0$) show extensive control over the ionization processes under experimentally feasible conditions.
\end{abstract}
\maketitle

Coherent control makes use of the quantum interference of different pathways to control the outcome of a physical process~\cite{Shapiro2012}. As a quantum interference effect it is exquisitely sensitive to the requirement of pathway indistinguishably. This condition  presents a major challenge for the 
control of bimolecular collisions, which requires entanglement of internal vibrational and external translational degrees of freedom for quantum interference to occur~\cite{Brumer1996,Brumer1999,Gong2003,Spanner2007}. However, there are two classes of processes for which these obstacles were conceptually overcome: the tetra-atomic reaction $\text{AB}+\text{AB}\rightarrow \text{A}_2+\text{B}_2$~\cite{Gong2003,Shapiro2012}, as well as Penning (PI) and Associative ionization (AI)~\cite{Arango2006a,Arango2006b}. The control of the PI and AI is extremely important in, e.g., the formation of Bose-Einstein condensates, since the ionization leads to losses in the trap~\cite{Shlyapnikov1994}. The PI and AI are also important in the spectroscopy of surfaces~\cite{Harada1997}, plasma chemistry~\cite{Roberge1982,Liu2010} or atmospheric studies~\cite{Falcinelli2015}. Similarly, control of $\text{AB}+\text{AB}$ would eliminate undesired chemical reactions in ultracold gases~\cite{Ospelkaus2010}. 

Although bond making has been demonstrated \cite{Levin2015,Levin2015a}, to date there has been no experimental demonstration of quantum interference controlled reactive collisional processes. However, experimental technology in the control and study of interatomic collisions has seen tremendous progress in recent years. In particular, advances in the control of ultracold atoms and molecules allows for a precise preparation and investigation of different collision scenarios~\cite{Vewinger2007,Vewinger2010,Shagam2013,Osterwalder2015} -- including PI at 10mK~\cite{henson2012}. In addition, the use of additional optical and static fields enables one to study the stereodynamics of the collisional processes~\cite{aoiz2015}. Very recently this ansatz has been used to obtain the branching ratio of PI and AI in the collision of Ne$^*(\tripletpp)$+Ar(\singlets)~\cite{Gordon2017,Gordon2018} Kr, Xe and N$_2$~\cite{Zou2018}, resolved for different initial incident angles. In light of these recent developments in experimental techniques and the absence of experimental demonstrations of collisional control, a reexamination of coherent control approaches in collisional processes is strongly motivated.

This letter provides a viable method for the coherent control of Penning  and Associative ionization as an example of the interference-based control of collisional processes. In addition to doing so we also resolve crucial symmetry issues that were inadvertently neglected in prior work~\cite{Arango2006b, Arango2006a}. Resolving these issues also offers considerable insight into conditions for interference in angular momentum controlled processes of interest here.

In Penning and Associative ionization (reviewed in~\cite{Siska1993}) two atoms A and B collide - with at least one (here B) being initially in a metastable state B$^*$ --, leading to ionization:
\begin{subequations}
\label{eq:penning_associative}
\begin{align}
  \text{A}+\text{B}^*&\xrightarrow{\text{PI}} \text{A}^++\text{B}+e^- \\
  \text{A}+\text{B}^*&\xrightarrow{\text{AI}}  (\text{AB})^++e^- \,.
\end{align}
\end{subequations}
In the Penning mechanism, the metastable species B$^*$ relaxes to its ground state while the other atom becomes ionized. In  AI, on the other hand, an ionic dimer (AB)$^+$ is formed. In both cases an electron is ejected. In general, PI and AI can decay to the same molecular channel, where a bound (AI) or a continuum state (PI) is formed.

The initial state of the colliding atoms may be written as a product~\cite{Arango2006b}
\begin{equation}
  \label{eq:initial_state}
  \ket{\psi}=\ket{\psi_{\text{A}}}\ket{\psi_{\text{B}}}
\end{equation}
where $\ket{\psi_{\text{A}}}$ and $\ket{\psi_{\text{B}}}$ are the initial states of the atom A and B, which, in the laboratory fixed frame (LFF), are
\begin{eqnarray}
  \label{eq:initial_a}
  \ket{\psi_{\text{A}}}&=&\exp\left(i\vec{k}_{\text{A}}\vec{r}_{\text{A}}\right)\sum_n a_n\ket{\phi^n_{\text{A}}}\\
  \label{eq:initial_b_star}
  \ket{\psi_{\text{B}}}&=&\exp\left(i\vec{k}_{\text{B}}\vec{r}_{\text{B}}\right)\sum_m b_m \ket{\phi^m_{\text{B}^*}}\,.
\end{eqnarray}
Here, $\vec{k}_{\text{X}}$ and $\vec{r}_{\text{X}}$ denote the linear momentum and position of atom X, and the sum on the right hand side is its electronic state. Note that the latter is a superposition of electronic eigenstates $\ket{\phi^i_{\text{X}}}$ with expansion coefficients $a_i$, $b_i$. For convenience, we introduce the center of mass $\vec{R}$ and the internuclear separation vector $\vec{r}$ as coordinates:
\begin{eqnarray}
  \label{eq:r_cm}
  \vec{R}&\equiv&\cfrac{m_{\text{A}} \vec{r}_{\text{A}}+m_{\text{B}} \vec{r}_{\text{B}}}{m_{\text{A}}+m_{\text{B}}}\\
  \label{eq:r}
  \vec{r}&\equiv&\vec{r}_{\text{B}}-\vec{r}_{\text{A}}\,,
\end{eqnarray}
where $m_{\text{X}}$ is the mass of species X.
The respective linear momenta are given by
\begin{eqnarray}
  \label{eq:k_cm}
  \vec{K}&\equiv& \vec{k}_{\text{A}}+\vec{k}_{\text{B}}\\
\label{eq:k}
  \vec{k}&\equiv& \cfrac{m_{\text{A}} \vec{k}_{\text{A}}-m_{\text{B}}\vec{k}_{\text{B}}}{m_{\text{A}}+m_{\text{B}}} \,.
\end{eqnarray}
Inserting expressions~\eqref{eq:r_cm}-\eqref{eq:k} into Eq.~\eqref{eq:initial_state} yields
\begin{equation}
  \label{eq:initial_state_cmf}
  \ket{\psi}=\exp\left(i\vec{K}\vec{R}+i\vec{k}\vec{r}\right)
  \sum_{n,m}a_nb_m\ket{\phi^n_{\text{A}}}\ket{\phi^m_{\text{B}^*}}.
\end{equation}

The total cross sections for PI and AI are given by
\begin{equation}
\label{eq:xs_pi_ai}
\sigma^\text{PI/AI}\left(\{c_S\}\right)=\sum\limits_{S,S'}c_S^*c_{S'}\sigma^\text{PI/AI}_{S,S'},
\end{equation}
where we have introduced $S\equiv(n,m)$, $c_S\equiv a_n b_m$, and $\ket{S}\equiv\ket{\phi^n_{\text{A}}}\ket{\phi^m_{\text{B}^*}}$. Here $\sigma_{S,S'}^\text{PI/AI}$ is the cross section for PI or AI associated with an initial state which is a superposition of $\ket{S}=\ket{\phi_\text{A}^n}\ket{\phi_{\text{B}^*}^m}$ and $\ket{S'}=\ket{\phi_\text{A}^{n'}}\ket{\phi_{\text{B}^*}^{m'}}$. The terms $\sigma^\text{PI/AI}_{S,S'}$ can be obtained by integrating the differential cross sections over continuum and bound final states~\cite{Taylor1972,Arango2006b}.

For the collision operator $\hat{W}$, the cross section into a specific exit channel $\ket{f}$ is given as~\cite{Shapiro2012}
\begin{align}
\sigma_f &= \left|\bra{f}\hat W \ket{\psi}\right|^2 \nonumber \\
&=\sum_{S,S'} c_S c_{S'}^* \bra{f}\hat W \ket{S} \bra{S'} \hat W \ket{f} \,.
\label{eq:interference}
\end{align}
From this expression we can compute the total cross section in Eq.~\eqref{eq:xs_pi_ai} by summing over all the final channels $\ket{f}$. Interference is manifest in the cross terms $S\neq S'$ in $\sigma$~\cite{Shapiro2012} with control affected via the coefficients $c_S$. For any symmetry that is conserved by $\hat W$, these cross terms are non-zero only if $\ket{f}$, $\ket{S}$, and $\ket{S'}$ are of the same symmetry~\footnote{Note that if the exit channel is not of a specific symmetry, but a superposition of different symmetries, it can couple initial states of different symmetry and non-zero cross terms results. However, when calculating the total cross section, and thus summing over all exit channels, the cross terms cancel}. For example, for the case of coherent control of collisional processes in the absence of an external, time-dependent field, the energy as a conserved quantity leads to the constraint that the initial state has to be a linear superposition of energetically \emph{degenerate} states. In past proposals for coherent control of the AI and PI cross sections, this constraint was taken into account by choosing a metastable state with non-zero angular momentum $J$, and using the degenerate states of different magnetic quantum number $M_J$ to construct the superposition state~\cite{Arango2006a,Arango2006b}.
However, as will be shown in the following, this may not be sufficient since we must also properly take into account the invariance of the collisional cross section under rotations in the LFF.

Specifically, in the absence of an external field, the collision process is invariant with respect to any rotation of the LFF. Consider then the collision between atoms A and B, and define the $Z$ axis of the LFF by the directional vector of the relative momentum $\vec{k}$ at $t\rightarrow -\infty$. Furthermore assume that the atoms A and B are initially in states with well-defined magnetic quantum numbers $M_{\text{A}}$ and $M_{\text{B}}$, which are the projection of each atom's electronic angular momentum on the $Z$-axis. The initial state then reads
  \begin{equation}
    \label{eq:psi_near_cm}
      \ket{\psi}=e^{\left(i\vec{K}\vec{R}+i k \vec{r}\hat{Z}\right)}\ket{M_{\text{A}}}\ket{M_{\text{B}}}.
  \end{equation}
 For convenience we apply a boost to the center of mass frame (CMF), where $\vec{K}=0$. In this frame, the wave function reads as
  \begin{equation}
    \label{eq:psi_near_cmf}
      \ket{\psi}=e^{i k \vec{r}\hat{Z}}\ket{M_{\text{A}}}\ket{M_{\text{B}}}.
  \end{equation}
Note that the total cross section is invariant with respect to a change  of the reference frame~\cite{Taylor1972}. Expression~\eqref{eq:psi_near_cmf} does not depend on the azimuthal angle, $\varphi$, that is, the orbital magnetic quantum number is $M_N=0$, so that, before the collision, the total magnetic quantum number of the A-B$^*$ system is $M_J=M_N+M_A+M_B=M_A+M_B$. Further, the Hamiltonian $H$ commutes with rotations around any axis of the LFF, and in particular around the $Z$-axis, i.e., $[H,\mathbf{J}_Z]=0$. Thus $M_J=M_A+M_B$ is a good quantum number, implying that the final state, in the asymptotic limit $t\rightarrow \infty$, either for PI or AI, is also characterized by the same $M_J$. Therefore, two different initial states $\ket{\psi}$ and $\ket{\psi'}$ can decay to the same final state with magnetic quantum numbers $M_A'$, $M_B'$ and hence interfere~\cite{Shapiro2012}, only if $M_A+M_B=M'_A+M'_B$.

This restriction is intimately limited to the initial superposition state rotated around the LFF. Consider a more general initial state
\begin{equation}
\ket{\psi}=e^{i\vec{k}\vec{r}}\ket{\psi_{\text{A}}}\ket{\psi_{\text{B}}} \,,
\end{equation}
where $\ket{\psi_{\text{A}}}=\left(a_1\ket{M^{(1)}_{\text{A}}}+a_2\ket{M^{(2)}_{\text{A}}}\right)$ and $\ket{\psi_{\text{B}}}=\left(b_1\ket{M^{(1)}_{\text{B}}}+b_2\ket{M^{(2)}_{\text{B}}}\right)$ are linear superpositions of magnetic quantum number states, with $M^{(1)}_{\text{A}} \neq M^{(2)}_{\text{A}}$ and $M^{(1)}_{\text{B}} \neq M^{(2)}_{\text{B}}$. Introducing the notation $c_{ij}\equiv a_i b_j$ and $\ket{S_{ij}}\equiv \ket{M^{(i)}_{\text{A}}}\ket{M^{(j)}_{\text{B}}}$, we can write
\begin{equation}
\ket{\psi}=e^{i\vec{k}\vec{r}}
\left( c_{11}\ket{S_{11}} + c_{12}\ket{S_{12}} + c_{21}\ket{S_{21}} + c_{22}\ket{S_{22}}\right)\,.
\label{eq:near_star_initial_state}
\end{equation}
Using Eq.~\eqref{eq:xs_pi_ai}, the ionization cross section is given as
\begin{align}
  \nonumber
  \sigma =&
    |c_{11}|^2\sigma_{11,11} + |c_{12}|^2\sigma_{12,12} + |c_{21}|^2 \sigma_{21,21} + |c_{22}|^2\sigma_{22,22} \nonumber\\
    +&2\text{Re}\{ c^*_{11}c_{12}\sigma_{11,12} + c^*_{11}c_{21}\sigma_{11,21}  + c^*_{11}c_{22}\sigma_{11,22}  \nonumber\\
    +& c^*_{12}c_{21}\sigma_{12,21} + c^*_{12}c_{22}\sigma_{12,22}  + c^*_{21}c_{22}\sigma_{21,22} \} \,.
\label{eq:sigma_initial_state_near_star}
\end{align}
Now, apply a rotation $R(\gamma)$ around the $Z$-axis of the LFF, which acts as $R(\gamma)e^{i\vec{k}\vec{r}}\ket{M_A}\ket{M_B}=e^{i\vec{k}\vec{r}}e^{i(M_A+M_B)\gamma}\ket{M_A}\ket{M_B}$. Using the notation $s_{ij}\equiv M^{(i)}_{\text{A}}+M^{(j)}_{\text{B}}$, the cross section for the rotated state, $\sigma'$, is
\begin{align}
  \nonumber
  \sigma' =&
    |c_{11}|^2\sigma_{11,11} + |c_{12}|^2\sigma_{12,12} + |c_{21}|^2 \sigma_{21,21} + |c_{22}|^2\sigma_{22,22} \nonumber\\
    +&2\text{Re}\{ c^*_{11}c_{12}\sigma_{11,12} e^{i(s_{12}-s_{11})\gamma}+ c^*_{11}c_{21}\sigma_{11,21} e^{i(s_{21}-s_{11})\gamma} \nonumber\\
    +& c^*_{11}c_{22}\sigma_{11,22} e^{i(s_{22}-s_{11})\gamma} + c^*_{12}c_{21}\sigma_{12,21} e^{i(s_{21}-s_{12})\gamma} \nonumber\\
    +& c^*_{12}c_{22}\sigma_{12,22} e^{i(s_{22}-s_{12})\gamma} + c^*_{21}c_{22}\sigma_{21,22} e^{i(s_{21}-s_{22})\gamma} \} \,.
\label{eq:sigma_initial_state_near_star_rotated}
\end{align}
Since a rotation of the LFF must not affect the total cross section, $\sigma=\sigma'$ has to hold for all angles $\gamma$. This can only be fulfilled if $\sigma_{ij,kl}=0$ for all $s_{ij} \neq s_{kl}$. Thus, the only non-zero cross terms $\sigma_{ij,kl}$ that can give rise to quantum interference are those with for $M_A^{(1)}+M_B^{(1)}=M_A^{(2)}+M_B^{(2)}$, which is the same condition on the conservation of the total magnetic quantum number stated above. Note that this condition is based on symmetry operations in the LFF, and is independent of the interaction between the atoms. It is therefore generic for any bimolecular collision using degenerate magnetic states for coherent control. This rotational symmetry perspective allows a simple principle to aid in formulating effective control scenarios. Specifically, any coherent control scenario must rely on initial states that are anisotropic with respect to rotations in the LFF.

Consider as an example PI of metastable Helium and Lithium~\cite{Merz1990a,Movre2000} (the AI cross section is here ignorable since HeLi$^+$ is very weakly bound):
\begin{equation}
\label{eq:hehe_scattering}
\text{He}^*\left(\triplets\right)+\text{Li}\left(\doublets\right)\rightarrow \text{He}(\singlets)+\text{Li}^+(\singlets)+e^- \,.
\end{equation}
Since both atoms have an orbital angular momentum of zero, one need only be  concerned with the electronic spin.
There are six spin states $\ket{S,M_S}_{\text{m}}$, where $S=\frac{1}{2}, \frac{3}{2}$ is the total spin of the diatomic complex and $M_S=S,S-1,...,-S$ is its projection onto the $Z$-axis (the direction of the relative momentum), and the subscript m denotes this as the molecular basis.
The states with a total spin of $S=\frac{3}{2}$ form the quartet $^{4}\Sigma$, the states with a total spin of $S=\frac{1}{2}$ the doublet $^{2}\Sigma$.

Only the doublet states are autoionizing~\cite{Movre2000}.
Thus it seems a logical choice to use a coherent superposition of the two doublet states for control:
\begin{equation}\label{eq.mol}
|\Psi\rangle = \frac{1}{\sqrt{2}} \ket{\frac{1}{2},\frac{1}{2}}_{\text{m}} + \frac{1}{\sqrt{2}} e^{i\beta} \ket{\frac{1}{2},-\frac{1}{2}}_{\text{m}} \,.
\end{equation}
where the cross section for PI from $\ket{\frac{1}{2},\frac{1}{2}}_{\text{m}}$ is $\sigma_+$ and from $\ket{\frac{1}{2},-\frac{1}{2}}_{\text{m}}$ is $\sigma_-$.
However, in accordance with the discussion above, varying $\beta$ will not provide control since the state is isotropic with respect to polarization in the LFF.

In addition, the state in Eq.~\eqref{eq.mol} is likely hard to prepare experimentally.
It is easier to create a superposition of the atomic states (both here and in the example below) by preparing the atoms prior to reaching the interaction zone, using established techniques such as electromagnetically induced transparency~\cite{Fleischhauer2005}, coherent population trapping~\cite{Arimondo1996}, or stimulated Raman adiabatic passage~\cite{Vitanov2017}.

Consider then an initial state, with both atoms in a superposition of atomic states
\begin{equation}
\label{eq.atom2}
\ket{\Psi} =\frac{1}{2} \bigg( \ket{1,1}_{\text{He}} + \ket{1,0}_{\text{He}} \bigg)\bigg(\ket{\frac{1}{2},\frac{1}{2}}_{\text{Li}} + e^{i\beta} \ket{\frac{1}{2},-\frac{1}{2}}_{\text{Li}} \bigg)\,.
\end{equation}
The phase $\beta$ for this state does not describe a rotation around the $Z$-axis, and the ionization cross section becomes
\begin{equation}
\label{eq.cross}
\sigma=\left(\frac{1}{4}-\frac{1}{3\sqrt{2}}\cos\beta\right)\sigma_+ + \frac{1}{12} \sigma_- \,.
\end{equation}
[This result is obtained by expressing the state~\eqref{eq.atom2} in the molecular basis.]
Thus, by properly taken the symmetry of the collision process into account, coherent control of this bimolecular reaction becomes possible, through variations in the angle $\beta$.

The above process only includes one ionization channel. 
As a second example that includes both AI and PI consider $\text{Ne}^*\left({^3\text{P}_2}\right)+\text{Ar}\left({^1\text{S}_0}\right)$.
This reaction was a prime focus in the proposed coherent control of a reactive bimolecular process~\cite{Arango2006a,Arango2006b},
where Ne$^*\left({^3\text{P}_2}\right)$ was initially in a superposition of Zeeman sublevels with projection $M_B^{(1)}$ and $M_B^{(2)}$ along the space fixed $Z$-axis and the Argon atom was in a single state with $M_A^{(1)}=0$. Control was said to be achieved by varying the amplitude and phases of the $M_B^{(1)}$, $M_B^{(2)}$ superposition state. However, the overall state is symmetric with respect to rotation around the LFF $Z$-axis. Hence, altering the phases and amplitudes of the $M_B^{(1)}$ and $M_B^{(2)}$ states on which the Ne$^*\left({^3\text{P}_2}\right)$ superposition is comprised can not, by the arguments above, result in control over the Ne$^*$-Ar ionization process; that is, the control components of the computations  in Ref.~\cite{Arango2006a,Arango2006b} were in error~\footnote{This error resulted from the wrong treatment of the branching ratios,  $W_{\Omega\rightarrow X}$, from an initial, $\Omega$, to a final channel, $X$. This assumption lead to erroneous autoionization widths $\Gamma_\Omega(r)$ only dependent on the initial channel. In this work, the autoionization widths depend on both the initial and the final channel by means of the branching ratios, $\Gamma_{\Omega\rightarrow X}(r)=\Gamma_\Omega(r) W_{\Omega\rightarrow X}$}. 

Consider, however, if the Ne$^*\left({^3\text{P}_2}\right)$ state is comprised of a superposition of states that are quantized with respect to the $X$-axis. Varying the relative phase between elements of the superposition would not correspond to a rotation in the LFF (in this basis the magnetic quantum number of the nuclear orbital angular momentum is not zero) and control by tuning the phase becomes possible.
Specifically, consider the initial state
\begin{equation}
\label{eq:near_0_2_x}
\ket{\Psi}=\ket{\text{Ar}}\otimes \left(a_0\ketx{20}+a_2\ketx{22}\right),
\end{equation}
where $a_0,a_2\in\mathbb{C}$ and $|a_0|^2+|a_2|^2=1$. Here $\ketx{2M}$ is the electronic state of Ne$^*\left(^3\text{P}_2\right)$ with $M$ being the magnetic quantum number along the $X$ axis of the laboratory frame.
In order to calculate the cross section, it is beneficial to use a $Z$-quantized basis, as this allows us to employ the rotating atom approximation~\cite{Mori1964}.
The $X$-quantized states $\ketx{2M}$ relate to the $Z$-quantized states $\ket{2M}$ by means of a $\pi/2$ rotation around the laboratory $Y$-axis~\cite{Zare1988}:
\begin{equation}
\label{eq:x_and_z_basis}
\ketx{2M}=\sum\limits_{M'=-2}^2d_{M',M}^2(\pi/2)\ket{2M},
\end{equation}
where $d_{M',M}^J(\theta)$ are the reduced Wigner matrix elements~\cite{Zare1988}. Substituting Eq.~\eqref{eq:x_and_z_basis} in Eq.~\eqref{eq:near_0_2_x} we obtain
\begin{widetext}
\begin{equation}
\label{eq:near_0_2_x_to_z}
\ket{\Psi}=\ket{\text{Ar}}\otimes \left[\left(\sqrt{\dfrac{3}{8}}a_0+\dfrac{a_2}{4}\right)(\ket{22}+\ket{2-2})+\dfrac{a_0}{2}(\ket{21}+\ket{2-1})+\left(\sqrt{\dfrac{3}{8}}a_2-\dfrac{a_0}{2}\right)\ket{20}\right]\,.
\end{equation}
\end{widetext}
Using the rotating atom approximation, the projection $M$ of the angular momentum in the LFF is related to the projection $\Omega$ of the electronic angular momentum along the internuclear axis of the Ne-Ar dimer as $\Omega=|M|$.
The cross sections $\sigma_{\Omega}$ for the three channels $\Omega=0,1,2$ can now be calculated as described in Refs.~\cite{Arango2006a,Arango2006b}.
For the reasons outlined above, there is no cross-term between states of different $M$, and therefore also between different $\Omega$, and thus the total cross section is given as
\begin{eqnarray}
\nonumber
\sigma_{\Psi}^\text{PI/AI}&=&2\left|\sqrt{\dfrac{3}{8}}a_0+\dfrac{a_2}{4}\right|^2\sigma_2^\text{PI/AI}+\dfrac{|a_2|^2}{2}\sigma_1^\text{PI/AI}\\
\label{eq:xs_a0_a2}
&&+\left|\sqrt{\dfrac{3}{8}}a_2-\dfrac{a_0}{2}\right|^2\sigma_0^\text{PI/AI}\,.
\end{eqnarray}

It is convenient to write $a_0$ and $a_2$ in terms of the Hopf coordinates in $\mathbb{C}^2$, such that $a_0=\sin\eta$ and $a_2=e^{i\xi}\cos\eta$ where $\eta\in[0,\pi]$ and $\xi\in[0,2\pi)$.
Note that thus $\eta$ describes incoherent control via the population, and $\xi$ coherent control via the relative phase.
\begin{figure}
\includegraphics[width=3.375 in]{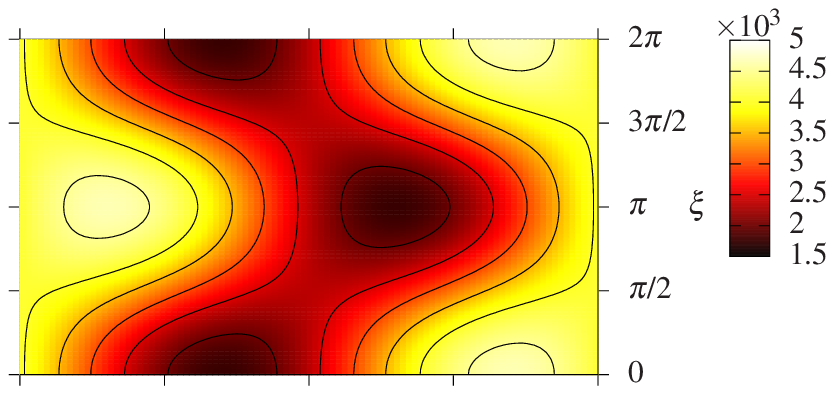}
\includegraphics[width=3.375 in]{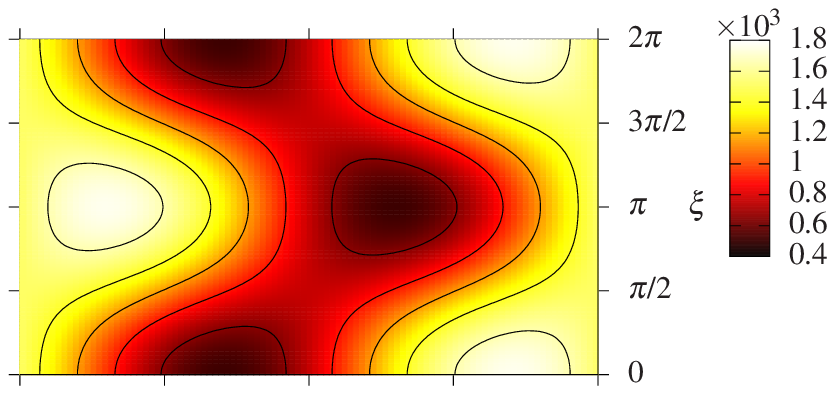}
\includegraphics[width=3.375 in]{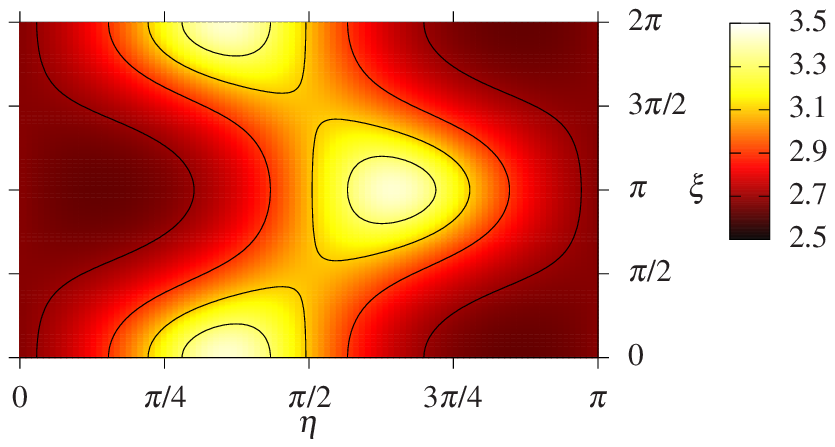}
\caption{\label{fig:near_0_2_xbasis_50mk} Cross sections for AI (upper panel), PI (middle panel), and the ratio of the two (lower panel) in the Ne$^*$-Ar system for the state~\eqref{eq:near_0_2_x}, as a function of  the control parameters $\eta$ and $\xi$.
The collision energy is 50 mK, and the cross sections are given in atomic units.
}
\end{figure}
Typical results at 50 mK are shown in Fig.~\ref{fig:near_0_2_xbasis_50mk} and the cross sections $\sigma_{\Omega}$ for each $\Omega$ are collected in Table~\ref{tab:sigma_omega}.
\begin{table}\centering
\caption{\label{tab:sigma_omega} Cross sections $\sigma_{\Omega}$ for Penning (PI) and associative (AI) ionization in the Ne$^*$-Ar system at T=50~mK, for the three different values of the projection $\Omega$ of the electronic angular momentum onto the internuclear axis. Cross sections are given in atomic units.}
\begin{ruledtabular}
  \begin{tabular}{ddd}
    \Omega &    \sigma^\text{PI} & \sigma^\text{AI}  \\
    \hline
     0 & 1997.95  & 5223.49 \\
     1 & 1447.30  & 3850.76 \\
     2 & 350.73  & 1392.43 \\
  \end{tabular}
\end{ruledtabular}
\end{table}
Evident is the extensive control of PI or AI as a function of $\eta$ and $\xi$. For example, the cross section for AI can be varied by means of $\eta$ and $\xi$
from 400 to 1700~a.u., \ie, by a factor of over 4. (This control is reduced to a factor of 2 if one  only tunes the relative phase $\xi$). Similar results are shown for control of $\sigma^\text{PI}$.
Interesting is that the functional dependence of $\sigma^\text{AI}$ and $\sigma^\text{PI}$ on $\eta$, $\xi$ are similar, so that control over the ratio $\sigma^\text{AI}/\sigma^\text{PI}$ is not as extensive (from 2.5 to 3.5). Hence, one can control total ionization by $\eta$ and $\xi$, an important result, for example, for suppression of ionization in a Bose-Einstein condensate~\cite{Shlyapnikov1994}.

In summary, we have introduced a viable means of controlling, via quantum interference effects, both Penning and Associative ionization in atomic collisions. Symmetry conditions were shown to provide insight into the selection of initial states for which control is possible. The axis of quantization along which superpositions of internal states is created was shown to be of particular importance in dictating the viability of control. In addition, significant control was demonstrated for two systems that are achievable with current scattering technologies.

\begin{acknowledgments}
This work was supported by the Natural Sciences and Engineering Research Council (NSERC) Canada.
\end{acknowledgments}


\begin{thebibliography}{35}%
\makeatletter
\providecommand \@ifxundefined [1]{%
 \@ifx{#1\undefined}
}%
\providecommand \@ifnum [1]{%
 \ifnum #1\expandafter \@firstoftwo
 \else \expandafter \@secondoftwo
 \fi
}%
\providecommand \@ifx [1]{%
 \ifx #1\expandafter \@firstoftwo
 \else \expandafter \@secondoftwo
 \fi
}%
\providecommand \natexlab [1]{#1}%
\providecommand \enquote  [1]{``#1''}%
\providecommand \bibnamefont  [1]{#1}%
\providecommand \bibfnamefont [1]{#1}%
\providecommand \citenamefont [1]{#1}%
\providecommand \href@noop [0]{\@secondoftwo}%
\providecommand \href [0]{\begingroup \@sanitize@url \@href}%
\providecommand \@href[1]{\@@startlink{#1}\@@href}%
\providecommand \@@href[1]{\endgroup#1\@@endlink}%
\providecommand \@sanitize@url [0]{\catcode `\\12\catcode `\$12\catcode
  `\&12\catcode `\#12\catcode `\^12\catcode `\_12\catcode `\%12\relax}%
\providecommand \@@startlink[1]{}%
\providecommand \@@endlink[0]{}%
\providecommand \url  [0]{\begingroup\@sanitize@url \@url }%
\providecommand \@url [1]{\endgroup\@href {#1}{\urlprefix }}%
\providecommand \urlprefix  [0]{URL }%
\providecommand \Eprint [0]{\href }%
\providecommand \doibase [0]{http://dx.doi.org/}%
\providecommand \selectlanguage [0]{\@gobble}%
\providecommand \bibinfo  [0]{\@secondoftwo}%
\providecommand \bibfield  [0]{\@secondoftwo}%
\providecommand \translation [1]{[#1]}%
\providecommand \BibitemOpen [0]{}%
\providecommand \bibitemStop [0]{}%
\providecommand \bibitemNoStop [0]{.\EOS\space}%
\providecommand \EOS [0]{\spacefactor3000\relax}%
\providecommand \BibitemShut  [1]{\csname bibitem#1\endcsname}%
\let\auto@bib@innerbib\@empty
\bibitem [{\citenamefont {Shapiro}\ and\ \citenamefont
  {Brumer}(2012)}]{Shapiro2012}%
  \BibitemOpen
  \bibfield  {author} {\bibinfo {author} {\bibfnamefont {M.}~\bibnamefont
  {Shapiro}}\ and\ \bibinfo {author} {\bibfnamefont {P.}~\bibnamefont
  {Brumer}},\ }\href@noop {} {\emph {\bibinfo {title} {{Quantum Control of
  Molecular Processes}}}}\ (\bibinfo  {publisher} {WILEY-VCH}, \bibinfo {address}{Weinheim},\ \bibinfo
  {year} {2012})\BibitemShut {NoStop}%
\bibitem [{\citenamefont {Shapiro}\ and\ \citenamefont
  {Brumer}(1996)}]{Brumer1996}%
  \BibitemOpen
  \bibfield  {author} {\bibinfo {author} {\bibfnamefont {M.}~\bibnamefont
  {Shapiro}}\ and\ \bibinfo {author} {\bibfnamefont {P.}~\bibnamefont
  {Brumer}},\ }\bibfield  {title} {\enquote {\bibinfo {title} {{Coherent
  control of collisional events: Bimolecular reactive scattering}},}\ }\href
  {http://www.ncbi.nlm.nih.gov/pubmed/10061988} {\bibfield  {journal} {\bibinfo
   {journal} {Phys. Rev. Lett.}\ }\textbf {\bibinfo {volume} {77}},\ \bibinfo
  {pages} {2574} (\bibinfo {year} {1996})}\BibitemShut {NoStop}%
\bibitem [{\citenamefont {Brumer}\ \emph {et~al.}(1999)\citenamefont {Brumer},
  \citenamefont {Abrashkevich},\ and\ \citenamefont {Shapiro}}]{Brumer1999}%
  \BibitemOpen
  \bibfield  {author} {\bibinfo {author} {\bibfnamefont {P.}\ \bibnamefont
  {Brumer}}, \bibinfo {author} {\bibfnamefont {A.}\ \bibnamefont
  {Abrashkevich}}, \ and\ \bibinfo {author} {\bibfnamefont {M.}\
  \bibnamefont {Shapiro}},\ }\bibfield  {title} {\enquote {\bibinfo {title}
  {{Laboratory conditions in the coherent control of reactive scattering}},}\
  }\href {\doibase 10.1039/A902135C} {\bibfield  {journal} {\bibinfo  {journal}
  {Faraday Discuss.}\ }\textbf {\bibinfo {volume} {113}},\ \bibinfo {pages}
  {291} (\bibinfo {year} {1999})}\BibitemShut {NoStop}%
\bibitem [{\citenamefont {Gong}\ \emph {et~al.}(2003)\citenamefont {Gong},
  \citenamefont {Shapiro},\ and\ \citenamefont {Brumer}}]{Gong2003}%
  \BibitemOpen
  \bibfield  {author} {\bibinfo {author} {\bibfnamefont {J.}\
  \bibnamefont {Gong}}, \bibinfo {author} {\bibfnamefont {M.}\ \bibnamefont
  {Shapiro}}, \ and\ \bibinfo {author} {\bibfnamefont {P.}\ \bibnamefont
  {Brumer}},\ }\bibfield  {title} {\enquote {\bibinfo {title}
  {{Entanglement-assisted coherent control in nonreactive diatom-diatom
  scattering}},}\ }\href {\doibase 10.1063/1.1535428} {\bibfield  {journal}
  {\bibinfo  {journal} {J. Chem. Phys.}\ }\textbf {\bibinfo {volume} {118}},\
  \bibinfo {pages} {2626} (\bibinfo {year} {2003})}\BibitemShut {NoStop}%
\bibitem [{\citenamefont {Spanner}\ and\ \citenamefont
  {Brumer}(2007)}]{Spanner2007}%
  \BibitemOpen
  \bibfield  {author} {\bibinfo {author} {\bibfnamefont {M.}\ \bibnamefont
  {Spanner}}\ and\ \bibinfo {author} {\bibfnamefont {P.}~\bibnamefont
  {Brumer}},\ }\bibfield  {title} {\enquote {\bibinfo {title} {{Entanglement
  and timing-based mechanisms in the coherent control of scattering
  processes}},}\ }\href {\doibase 10.1103/PhysRevA.76.013408} {\bibfield
  {journal} {\bibinfo  {journal} {Phys. Rev. A}\ }\textbf {\bibinfo {volume}
  {76}},\ \bibinfo {pages} {013408} (\bibinfo {year} {2007})}\BibitemShut
  {NoStop}%
\bibitem [{\citenamefont {Arango}\ \emph
  {et~al.}(2006{\natexlab{a}})\citenamefont {Arango}, \citenamefont {Shapiro},\
  and\ \citenamefont {Brumer}}]{Arango2006a}%
  \BibitemOpen
  \bibfield  {author} {\bibinfo {author} {\bibfnamefont {C.~A.}\ \bibnamefont
  {Arango}}, \bibinfo {author} {\bibfnamefont {M.}~\bibnamefont {Shapiro}}, \
  and\ \bibinfo {author} {\bibfnamefont {P.}~\bibnamefont {Brumer}},\
  }\bibfield  {title} {\enquote {\bibinfo {title} {{Cold atomic collisions:
  Coherent control of penning and associative ionization}},}\ }\href {\doibase
  10.1103/PhysRevLett.97.193202} {\bibfield  {journal} {\bibinfo  {journal}
  {Phys. Rev. Lett.}\ }\textbf {\bibinfo {volume} {97}},\ \bibinfo {pages}
  {193202} (\bibinfo {year} {2006}{\natexlab{a}})}\BibitemShut {NoStop}%
\bibitem [{\citenamefont {Arango}\ \emph
  {et~al.}(2006{\natexlab{b}})\citenamefont {Arango}, \citenamefont {Shapiro},\
  and\ \citenamefont {Brumer}}]{Arango2006b}%
  \BibitemOpen
  \bibfield  {author} {\bibinfo {author} {\bibfnamefont {C.~A.}\ \bibnamefont
  {Arango}}, \bibinfo {author} {\bibfnamefont {M.}~\bibnamefont {Shapiro}}, \
  and\ \bibinfo {author} {\bibfnamefont {P.}~\bibnamefont {Brumer}},\
  }\bibfield  {title} {\enquote {\bibinfo {title} {{Coherent control of
  collision processes: Penning versus associative ionization}},}\ }\href
  {\doibase 10.1063/1.2336430} {\bibfield  {journal} {\bibinfo  {journal} {J.
  Chem. Phys.}\ }\textbf {\bibinfo {volume} {125}},\ \bibinfo {pages} {094315}
  (\bibinfo {year} {2006}{\natexlab{b}})}\BibitemShut {NoStop}%
\bibitem [{\citenamefont {Shlyapnikov}\ \emph {et~al.}(1994)\citenamefont
  {Shlyapnikov}, \citenamefont {Walraven}, \citenamefont {Rahmanov},\ and\
  \citenamefont {Reynolds}}]{Shlyapnikov1994}%
  \BibitemOpen
  \bibfield  {author} {\bibinfo {author} {\bibfnamefont {G.~V.}\ \bibnamefont
  {Shlyapnikov}}, \bibinfo {author} {\bibfnamefont {J.~T. M.}\ \bibnamefont
  {Walraven}}, \bibinfo {author} {\bibfnamefont {U.~M.}\ \bibnamefont
  {Rahmanov}}, \ and\ \bibinfo {author} {\bibfnamefont {M.~W.}\ \bibnamefont
  {Reynolds}},\ }\bibfield  {title} {\enquote {\bibinfo {title} {{Decay
  kinetics and bose condensation in a gas of spin-polarized triplet helium}},}\
  }\href {\doibase 10.1103/PhysRevLett.73.3247} {\bibfield  {journal} {\bibinfo
   {journal} {Phys. Rev. Lett.}\ }\textbf {\bibinfo {volume} {73}},\ \bibinfo
  {pages} {3247} (\bibinfo {year} {1994})}\BibitemShut {NoStop}%
\bibitem [{\citenamefont {Harada}\ \emph {et~al.}(1997)\citenamefont {Harada},
  \citenamefont {Masuda},\ and\ \citenamefont {Ozaki}}]{Harada1997}%
  \BibitemOpen
  \bibfield  {author} {\bibinfo {author} {\bibfnamefont {Y.}\ \bibnamefont
  {Harada}}, \bibinfo {author} {\bibfnamefont {S.}\ \bibnamefont
  {Masuda}}, \ and\ \bibinfo {author} {\bibfnamefont {H.}\ \bibnamefont
  {Ozaki}},\ }\bibfield  {title} {\enquote {\bibinfo {title} {{Electron
  Spectroscopy Using Metastable Atoms as Probes for Solid Surfaces}},}\ }\href
  {\doibase 10.1021/cr940315v} {\bibfield  {journal} {\bibinfo  {journal}
  {Chem. Rev.}\ }\textbf {\bibinfo {volume} {97}},\ \bibinfo {pages}
  {1897} (\bibinfo {year} {1997})}\BibitemShut {NoStop}%
\bibitem [{\citenamefont {Roberge}\ and\ \citenamefont
  {Dalgarno}(1982)}]{Roberge1982}%
  \BibitemOpen
  \bibfield  {author} {\bibinfo {author} {\bibfnamefont {W.}~\bibnamefont
  {Roberge}}\ and\ \bibinfo {author} {\bibfnamefont {A.}~\bibnamefont
  {Dalgarno}},\ }\bibfield  {title} {\enquote {\bibinfo {title} {{The Formation
  and Destruction of HeH$^+$ in astrophysical plasmas}},}\ }\href@noop {}
  {\bibfield  {journal} {\bibinfo  {journal} {Astrophys. J.}\ }\textbf
  {\bibinfo {volume} {255}},\ \bibinfo {pages} {489} (\bibinfo {year}
  {1982})}\BibitemShut {NoStop}%
\bibitem [{\citenamefont {Liu}\ \emph {et~al.}(2010)\citenamefont {Liu},
  \citenamefont {Bruggeman}, \citenamefont {Iza}, \citenamefont {Rong},\ and\
  \citenamefont {Kong}}]{Liu2010}%
  \BibitemOpen
  \bibfield  {author} {\bibinfo {author} {\bibfnamefont {D. X.}\ \bibnamefont
  {Liu}}, \bibinfo {author} {\bibfnamefont {P.}~\bibnamefont {Bruggeman}},
  \bibinfo {author} {\bibfnamefont {F.}~\bibnamefont {Iza}}, \bibinfo {author}
  {\bibfnamefont {M. Z.}\ \bibnamefont {Rong}}, \ and\ \bibinfo {author}
  {\bibfnamefont {M. G.}\ \bibnamefont {Kong}},\ }\bibfield  {title} {\enquote
  {\bibinfo {title} {{Global model of low-temperature atmospheric-pressure He +
  H$_2$O plasmas}},}\ }\href {http://stacks.iop.org/0963-0252/19/i=2/a=025018}
  {\bibfield  {journal} {\bibinfo  {journal} {Plasma Sources Sci. Technol.}\
  }\textbf {\bibinfo {volume} {19}},\ \bibinfo {pages} {25018} (\bibinfo {year}
  {2010})}\BibitemShut {NoStop}%
\bibitem [{\citenamefont {Falcinelli}\ \emph {et~al.}(2015)\citenamefont
  {Falcinelli}, \citenamefont {Pirani},\ and\ \citenamefont
  {Vecchiocattivi}}]{Falcinelli2015}%
  \BibitemOpen
  \bibfield  {author} {\bibinfo {author} {\bibfnamefont {S.}\ \bibnamefont
  {Falcinelli}}, \bibinfo {author} {\bibfnamefont {F.}\ \bibnamefont
  {Pirani}}, \ and\ \bibinfo {author} {\bibfnamefont {F.}\ \bibnamefont
  {Vecchiocattivi}},\ }\bibfield  {title} {\enquote {\bibinfo {title} {{The
  possible role of penning ionization processes in planetary atmospheres}},}\
  }\href {\doibase 10.3390/atmos6030299} {\bibfield  {journal} {\bibinfo
  {journal} {Atmosphere}\ }\textbf {\bibinfo {volume} {6}},\ \bibinfo
  {pages} {299} (\bibinfo {year} {2015})}\BibitemShut {NoStop}%
\bibitem [{\citenamefont {Ospelkaus}\ \emph {et~al.}(2010)\citenamefont
  {Ospelkaus}, \citenamefont {Ni}, \citenamefont {Wang}, \citenamefont
  {de~Miranda}, \citenamefont {Neyenhuis}, \citenamefont
  {Qu{\'{e}}m{\'{e}}ner}, \citenamefont {Julienne}, \citenamefont {Bohn},
  \citenamefont {Jin},\ and\ \citenamefont {Ye}}]{Ospelkaus2010}%
  \BibitemOpen
  \bibfield  {author} {\bibinfo {author} {\bibfnamefont {S.}~\bibnamefont
  {Ospelkaus}}, \bibinfo {author} {\bibfnamefont {K. K.}\ \bibnamefont {Ni}},
  \bibinfo {author} {\bibfnamefont {D.}~\bibnamefont {Wang}}, \bibinfo {author}
  {\bibfnamefont {M. H. G.}\ \bibnamefont {de~Miranda}}, \bibinfo {author}
  {\bibfnamefont {B.}~\bibnamefont {Neyenhuis}}, \bibinfo {author}
  {\bibfnamefont {G.}~\bibnamefont {Qu{\'{e}}m{\'{e}}ner}}, \bibinfo {author}
  {\bibfnamefont {P. S.}\ \bibnamefont {Julienne}}, \bibinfo {author}
  {\bibfnamefont {J. L.}\ \bibnamefont {Bohn}}, \bibinfo {author} {\bibfnamefont
  {D. S.}\ \bibnamefont {Jin}}, \ and\ \bibinfo {author} {\bibfnamefont
  {J.}~\bibnamefont {Ye}},\ }\bibfield  {title} {\enquote {\bibinfo {title}
  {{Quantum-state controlled chemical reactions of ultracold potassium-rubidium
  molecules}},}\ }\href {\doibase 10.1126/science.1184121} {\bibfield
  {journal} {\bibinfo  {journal} {Science}\ }\textbf {\bibinfo {volume}
  {327}},\ \bibinfo {pages} {853} (\bibinfo {year} {2010})}\BibitemShut
  {NoStop}%
\bibitem [{\citenamefont {Levin}\ \emph
  {et~al.}(2015{\natexlab{a}})\citenamefont {Levin}, \citenamefont
  {Skomorowski}, \citenamefont {Rybak}, \citenamefont {Kosloff}, \citenamefont
  {Koch},\ and\ \citenamefont {Amitay}}]{Levin2015}%
  \BibitemOpen
  \bibfield  {author} {\bibinfo {author} {\bibfnamefont {L.}\ \bibnamefont
  {Levin}}, \bibinfo {author} {\bibfnamefont {W.}\ \bibnamefont
  {Skomorowski}}, \bibinfo {author} {\bibfnamefont {Leonid}\ \bibnamefont
  {Rybak}}, \bibinfo {author} {\bibfnamefont {R.}\ \bibnamefont {Kosloff}},
  \bibinfo {author} {\bibfnamefont {C. P.}\ \bibnamefont {Koch}}, \ and\
  \bibinfo {author} {\bibfnamefont {Z.}\ \bibnamefont {Amitay}},\ }\bibfield
   {title} {\enquote {\bibinfo {title} {{Coherent Control of Bond Making}},}\
  }\href {\doibase 10.1103/PhysRevLett.114.233003} {\bibfield  {journal}
  {\bibinfo  {journal} {Phys. Rev. Lett.}\ }\textbf {\bibinfo {volume} {114}},\
  \bibinfo {pages} {233003} (\bibinfo {year} {2015}{\natexlab{a}})}\BibitemShut
  {NoStop}%
\bibitem [{\citenamefont {Levin}\ \emph
  {et~al.}(2015{\natexlab{b}})\citenamefont {Levin}, \citenamefont
  {Skomorowski}, \citenamefont {Kosloff}, \citenamefont {Koch},\ and\
  \citenamefont {Amitay}}]{Levin2015a}%
  \BibitemOpen
  \bibfield  {author} {\bibinfo {author} {\bibfnamefont {L.}\ \bibnamefont
  {Levin}}, \bibinfo {author} {\bibfnamefont {W.}\ \bibnamefont
  {Skomorowski}}, \bibinfo {author} {\bibfnamefont {R.}\ \bibnamefont
  {Kosloff}}, \bibinfo {author} {\bibfnamefont {C. P.}\ \bibnamefont
  {Koch}}, \ and\ \bibinfo {author} {\bibfnamefont {Z.}\ \bibnamefont
  {Amitay}},\ }\bibfield  {title} {\enquote {\bibinfo {title} {{Coherent
  control of bond making: the performance of rationally phase-shaped
  femtosecond laser pulses}},}\ }\href {\doibase
  10.1088/0953-4075/48/18/184004} {\bibfield  {journal} {\bibinfo  {journal}
  {J. Phys. B At. Mol. Opt. Phys.}\ }\textbf {\bibinfo {volume} {48}},\
  \bibinfo {pages} {184004} (\bibinfo {year} {2015}{\natexlab{b}})}\BibitemShut
  {NoStop}%
\bibitem [{\citenamefont {Vewinger}\ \emph {et~al.}(2007)\citenamefont
  {Vewinger}, \citenamefont {Heinz}, \citenamefont {Schneider}, \citenamefont
  {Barthel},\ and\ \citenamefont {Bergmann}}]{Vewinger2007}%
  \BibitemOpen
  \bibfield  {author} {\bibinfo {author} {\bibfnamefont {F.}~\bibnamefont
  {Vewinger}}, \bibinfo {author} {\bibfnamefont {M.}~\bibnamefont {Heinz}},
  \bibinfo {author} {\bibfnamefont {U.}~\bibnamefont {Schneider}}, \bibinfo
  {author} {\bibfnamefont {C.}~\bibnamefont {Barthel}}, \ and\ \bibinfo
  {author} {\bibfnamefont {K.}~\bibnamefont {Bergmann}},\ }\bibfield  {title}
  {\enquote {\bibinfo {title} {{Amplitude and phase control of a coherent
  superposition of degenerate states. II. Experiment}},}\ }\href {\doibase
  10.1103/PhysRevA.75.043407} {\bibfield  {journal} {\bibinfo  {journal} {Phys.
  Rev. A}\ }\textbf {\bibinfo {volume} {75}},\ \bibinfo {pages} {043407}
  (\bibinfo {year} {2007})}\BibitemShut {NoStop}%
\bibitem [{\citenamefont {Vewinger}\ \emph {et~al.}(2010)\citenamefont
  {Vewinger}, \citenamefont {Shore},\ and\ \citenamefont
  {Bergmann}}]{Vewinger2010}%
  \BibitemOpen
  \bibfield  {author} {\bibinfo {author} {\bibfnamefont {F.}\ \bibnamefont
  {Vewinger}}, \bibinfo {author} {\bibfnamefont {B.~W.}\ \bibnamefont
  {Shore}}, \ and\ \bibinfo {author} {\bibfnamefont {K.}\ \bibnamefont
  {Bergmann}},\ }\bibfield  {title} {\enquote {\bibinfo {title}
  {{Superpositions of Degenerate Quantum States: Preparation and Detection in
  Atomic Beams}},}\ }\href {\doibase 10.1016/S1049-250X(10)05808-8} {\bibfield
  {journal} {\bibinfo  {journal} {Adv. At. Mol. Opt. Phys.}\ }\textbf {\bibinfo
  {volume} {58}},\ \bibinfo {pages} {113} (\bibinfo {year}
  {2010})}\BibitemShut {NoStop}%
\bibitem [{\citenamefont {Shagam}\ and\ \citenamefont
  {Narevicius}(2013)}]{Shagam2013}%
  \BibitemOpen
  \bibfield  {author} {\bibinfo {author} {\bibfnamefont {Y.}~\bibnamefont
  {Shagam}}\ and\ \bibinfo {author} {\bibfnamefont {E.}~\bibnamefont
  {Narevicius}},\ }\bibfield  {title} {\enquote {\bibinfo {title} {{Sub-Kelvin
  collision temperatures in merged neutral beams by correlation in
  phase-space}},}\ }\href {\doibase 10.1021/jp4045868} {\bibfield  {journal}
  {\bibinfo  {journal} {J. Phys. Chem. C}\ }\textbf {\bibinfo {volume} {117}},\
  \bibinfo {pages} {22454} (\bibinfo {year} {2013})}\BibitemShut
  {NoStop}%
\bibitem [{\citenamefont {Osterwalder}(2015)}]{Osterwalder2015}%
  \BibitemOpen
  \bibfield  {author} {\bibinfo {author} {\bibfnamefont {A.}\ \bibnamefont
  {Osterwalder}},\ }\bibfield  {title} {\enquote {\bibinfo {title} {{Merged
  neutral beams}},}\ }\href {\doibase 10.1140/epjti/s40485-015-0022-x}
  {\bibfield  {journal} {\bibinfo  {journal} {EPJ Tech. Instrum.}\ }\textbf
  {\bibinfo {volume} {2}},\ \bibinfo {pages} {10} (\bibinfo {year}
  {2015})}\BibitemShut {NoStop}%
\bibitem [{\citenamefont {Henson}\ \emph {et~al.}(2012)\citenamefont {Henson},
  \citenamefont {Gersten}, \citenamefont {Shagam}, \citenamefont {Narevicius},\
  and\ \citenamefont {Narevicius}}]{henson2012}%
  \BibitemOpen
  \bibfield  {author} {\bibinfo {author} {\bibfnamefont {A.~B.}\ \bibnamefont
  {Henson}}, \bibinfo {author} {\bibfnamefont {S.}~\bibnamefont {Gersten}},
  \bibinfo {author} {\bibfnamefont {Y.}~\bibnamefont {Shagam}}, \bibinfo
  {author} {\bibfnamefont {J.}~\bibnamefont {Narevicius}}, \ and\ \bibinfo
  {author} {\bibfnamefont {E.}~\bibnamefont {Narevicius}},\ }\bibfield  {title}
  {\enquote {\bibinfo {title} {{Observation of Resonances in Penning Ionization
  Reactions at Sub-Kelvin Temperatures in Merged Beams}},}\ }\href {\doibase
  10.1126/science.1229141} {\bibfield  {journal} {\bibinfo  {journal}
  {Science}\ }\textbf {\bibinfo {volume} {338}},\ \bibinfo {pages} {234}
  (\bibinfo {year} {2012})}\BibitemShut {NoStop}%
\bibitem [{\citenamefont {Aoiz}\ \emph {et~al.}(2015)\citenamefont {Aoiz},
  \citenamefont {Brouard}, \citenamefont {Gordon}, \citenamefont {Nichols},
  \citenamefont {Stolte},\ and\ \citenamefont {Walpole}}]{aoiz2015}%
  \BibitemOpen
  \bibfield  {author} {\bibinfo {author} {\bibfnamefont {F.~J.}\ \bibnamefont
  {Aoiz}}, \bibinfo {author} {\bibfnamefont {M.}~\bibnamefont {Brouard}},
  \bibinfo {author} {\bibfnamefont {S.~D.~S.}\ \bibnamefont {Gordon}}, \bibinfo
  {author} {\bibfnamefont {B.}~\bibnamefont {Nichols}}, \bibinfo {author}
  {\bibfnamefont {S.}~\bibnamefont {Stolte}}, \ and\ \bibinfo {author}
  {\bibfnamefont {V.}~\bibnamefont {Walpole}},\ }\bibfield{title} {\enquote {\bibinfo {title} {{A new perspective: Imaging
  the stereochemistry of molecular collisions}},}\ }\href {\doibase
  10.1039/c5cp03273c}  {\bibfield  {journal} {\bibinfo
  {journal} {Phys. Chem. Chem. Phys.}\ }\textbf {\bibinfo {volume} {17}},\ \bibinfo
  {pages} {30210} (\bibinfo {year} {2015})} \BibitemShut {NoStop}%
\bibitem [{\citenamefont {Gordon}\ \emph {et~al.}(2017)\citenamefont {Gordon},
  \citenamefont {Zou}, \citenamefont {Tanteri}, \citenamefont {Jankunas},\ and\
  \citenamefont {Osterwalder}}]{Gordon2017}%
  \BibitemOpen
  \bibfield  {author} {\bibinfo {author} {\bibfnamefont {S. D. S.}\
  \bibnamefont {Gordon}}, \bibinfo {author} {\bibfnamefont {J.}\
  \bibnamefont {Zou}}, \bibinfo {author} {\bibfnamefont {S.}\ \bibnamefont
  {Tanteri}}, \bibinfo {author} {\bibfnamefont {J.}\ \bibnamefont
  {Jankunas}}, \ and\ \bibinfo {author} {\bibfnamefont {A.}\ \bibnamefont
  {Osterwalder}},\ }\bibfield  {title} {\enquote {\bibinfo {title} {{Energy
  Dependent Stereodynamics of the
  $\mathrm{Ne}({^{3}\mathrm{P}}_{2})+\mathrm{Ar}$ Reaction}},}\ }\href
  {\doibase 10.1103/PhysRevLett.119.053001} {\bibfield  {journal} {\bibinfo
  {journal} {Phys. Rev. Lett.}\ }\textbf {\bibinfo {volume} {119}},\ \bibinfo
  {pages} {053001} (\bibinfo {year} {2017})}\BibitemShut {NoStop}%
\bibitem [{\citenamefont {Gordon}\ \emph {et~al.}(2018)\citenamefont {Gordon},
  \citenamefont {Omiste}, \citenamefont {Zou}, \citenamefont {Tanteri},
  \citenamefont {Jankunas}, \citenamefont {Brumer},\ and\ \citenamefont
  {Osterwalder}}]{Gordon2018}%
  \BibitemOpen
  \bibfield  {author} {\bibinfo {author} {\bibfnamefont {S.~D.~S.}\
  \bibnamefont {Gordon}}, \bibinfo {author} {\bibfnamefont {J.~J.}\
  \bibnamefont {Omiste}}, \bibinfo {author} {\bibfnamefont {J.}~\bibnamefont
  {Zou}}, \bibinfo {author} {\bibfnamefont {S.}~\bibnamefont {Tanteri}},
  \bibinfo {author} {\bibfnamefont {J.}~\bibnamefont {Jankunas}}, \bibinfo
  {author} {\bibfnamefont {P.}~\bibnamefont {Brumer}}, \ and\ \bibinfo {author}
  {\bibfnamefont {A.}~\bibnamefont {Osterwalder}},\ }\bibfield  {title}
  {\enquote {\bibinfo {title} {{Quantum State Controlled Channel Branching in
  Cold Ne$(^3\text{P}_2)$+Ar Chemi-Ionisation}},}\ }\href@noop {} {\bibfield
  {journal} {\bibinfo  {journal} {(unpublished)}\ } (\bibinfo {year}
  {2018})}\BibitemShut {NoStop}%
\bibitem [{\citenamefont {Zou}\ \emph {et~al.}(2018)\citenamefont {Zou},
  \citenamefont {Gordon}, \citenamefont {Tanteri},\ and\ \citenamefont
  {Osterwalder}}]{Zou2018}%
  \BibitemOpen
  \bibfield  {author} {\bibinfo {author} {\bibfnamefont {J.}\ \bibnamefont
  {Zou}}, \bibinfo {author} {\bibfnamefont {S. D. S.}\ \bibnamefont
  {Gordon}}, \bibinfo {author} {\bibfnamefont {S.}\ \bibnamefont
  {Tanteri}}, \ and\ \bibinfo {author} {\bibfnamefont {A.}\ \bibnamefont
  {Osterwalder}},\ }\bibfield  {title} {\enquote {\bibinfo {title}
  {{Stereodynamics of Ne($^3\text{P}_2$) reacting with Ar, Kr, Xe, and
  N$_2$}},}\ }\href {\doibase 10.1063/1.5026952} {\bibfield  {journal}
  {\bibinfo  {journal} {J. Chem. Phys.}\ }\textbf {\bibinfo {volume} {148}},\
  \bibinfo {pages} {164310} (\bibinfo {year} {2018})}\BibitemShut {NoStop}%
\bibitem [{\citenamefont {Siska}(1993)}]{Siska1993}%
  \BibitemOpen
  \bibfield  {author} {\bibinfo {author} {\bibfnamefont {P.~E.}\ \bibnamefont
  {Siska}},\ }\bibfield  {title} {\enquote {\bibinfo {title} {{Molecular-beam
  studies of Penning ionization}},}\ }\href {\doibase
  10.1103/RevModPhys.65.337} {\bibfield  {journal} {\bibinfo  {journal} {Rev.
  Mod. Phys.}\ }\textbf {\bibinfo {volume} {65}},\ \bibinfo {pages} {337}
  (\bibinfo {year} {1993})}\BibitemShut {NoStop}%
\bibitem [{\citenamefont {Taylor}(1972)}]{Taylor1972}%
  \BibitemOpen
  \bibfield  {author} {\bibinfo {author} {\bibfnamefont {J.~R.}\ \bibnamefont
  {Taylor}},\ }\href@noop {} {\emph {\bibinfo {title} {{Scattering Theory: The
  Quantum Theory of Nonrelativistic Collisions}}}}\ (\bibinfo{publisher}{John Wiley \& Sons}, \bibinfo  {address} {New York},\ \bibinfo {year} {1972})\BibitemShut {NoStop}%
\bibitem [{Note1()}]{Note1}%
  \BibitemOpen
  \bibinfo {note} {Note that if the exit channel is not of a specific symmetry,
  but a superposition of different symmetries, it can couple initial states of
  different symmetry and non-zero cross terms results. However, when
  calculating the total cross section, and thus summing over all exit channels,
  the cross terms cancel}\BibitemShut {NoStop}%
\bibitem [{\citenamefont {Merz}\ \emph {et~al.}(1990)\citenamefont {Merz},
  \citenamefont {M{\"{u}}ller}, \citenamefont {Ruf}, \citenamefont {Hotop},
  \citenamefont {Mayer},\ and\ \citenamefont {Movre}}]{Merz1990a}%
  \BibitemOpen
  \bibfield  {author} {\bibinfo {author} {\bibfnamefont {A.}~\bibnamefont
  {Merz}}, \bibinfo {author} {\bibfnamefont {M.W.}\ \bibnamefont
  {M{\"{u}}ller}}, \bibinfo {author} {\bibfnamefont {M.-W.}\ \bibnamefont
  {Ruf}}, \bibinfo {author} {\bibfnamefont {H.}~\bibnamefont {Hotop}}, \bibinfo
  {author} {\bibfnamefont {W.}~\bibnamefont {Mayer}}, \ and\ \bibinfo {author}
  {\bibfnamefont {M.}~\bibnamefont {Movre}},\ }\bibfield  {title} {\enquote
  {\bibinfo {title} {{Experimental and theoretical studies of simple attractive
  Penning ionization systems}},}\ }\href {\doibase
  10.1016/0301-0104(90)89117-9} {\bibfield  {journal} {\bibinfo  {journal}
  {Chem. Phys.}\ }\textbf {\bibinfo {volume} {145}},\ \bibinfo {pages}
  {219} (\bibinfo {year} {1990})}\BibitemShut {NoStop}%
\bibitem [{\citenamefont {Movre}\ \emph {et~al.}(2000)\citenamefont {Movre},
  \citenamefont {Thiel},\ and\ \citenamefont {Meyer}}]{Movre2000}%
  \BibitemOpen
  \bibfield  {author} {\bibinfo {author} {\bibfnamefont {M.}~\bibnamefont
  {Movre}}, \bibinfo {author} {\bibfnamefont {L.}~\bibnamefont {Thiel}}, \ and\
  \bibinfo {author} {\bibfnamefont {W.}~\bibnamefont {Meyer}},\ }\bibfield
  {title} {\enquote {\bibinfo {title} {{Theoretical investigation of the
  autoionization process in molecular collision complexes:
  He*(2$^3$S)+Li(2$^2$S)$\rightarrow$ He+Li$^+$+e${}^-$}},}\ }\href {\doibase
  10.1063/1.481935} {\bibfield  {journal} {\bibinfo  {journal} {J. Chem.
  Phys.}\ }\textbf {\bibinfo {volume} {113}},\ \bibinfo {pages} {1484}
  (\bibinfo {year} {2000})}\BibitemShut {NoStop}%
\bibitem [{\citenamefont {Fleischhauer}\ \emph {et~al.}(2005)\citenamefont
  {Fleischhauer}, \citenamefont {Imamoglu},\ and\ \citenamefont
  {Marangos}}]{Fleischhauer2005}%
  \BibitemOpen
  \bibfield  {author} {\bibinfo {author} {\bibfnamefont {M.}\ \bibnamefont
  {Fleischhauer}}, \bibinfo {author} {\bibfnamefont {A.}\ \bibnamefont
  {Imamoglu}}, \ and\ \bibinfo {author} {\bibfnamefont {J. P.}\
  \bibnamefont {Marangos}},\ }\bibfield  {title} {\enquote {\bibinfo {title}
  {{Electromagnetically induced transparency: Optics in coherent media}},}\
  }\href {\doibase 10.1103/RevModPhys.77.633} {\bibfield  {journal} {\bibinfo
  {journal} {Rev. Mod. Phys.}\ }\textbf {\bibinfo {volume} {77}},\ \bibinfo
  {pages} {633} (\bibinfo {year} {2005})}\BibitemShut {NoStop}%
\bibitem [{\citenamefont {Arimondo}(1996)}]{Arimondo1996}%
  \BibitemOpen
  \bibfield  {author} {\bibinfo {author} {\bibfnamefont {E.}~\bibnamefont
  {Arimondo}},\ }\bibfield  {title} {\enquote {\bibinfo {title} {V coherent
  population trapping in laser spectroscopy},}\ }in\ \href {\doibase
  https://doi.org/10.1016/S0079-6638(08)70531-6} {\emph {\bibinfo {booktitle}
  {Progress in Optics}}},\ Vol.~\bibinfo {volume} {35},\ \bibinfo {editor}
  {edited by\ \bibinfo {editor} {\bibfnamefont {E.}~\bibnamefont {Wolf}}}\
  (\bibinfo  {publisher} {Elsevier},\ \bibinfo {year} {1996})\ \bibinfo
  {pages} {257 -- 354}\BibitemShut {NoStop}%
\bibitem [{\citenamefont {Vitanov}\ \emph {et~al.}(2017)\citenamefont
  {Vitanov}, \citenamefont {Rangelov}, \citenamefont {Shore},\ and\
  \citenamefont {Bergmann}}]{Vitanov2017}%
  \BibitemOpen
  \bibfield  {author} {\bibinfo {author} {\bibfnamefont {N.~V.}\
  \bibnamefont {Vitanov}}, \bibinfo {author} {\bibfnamefont {A.~A.}\
  \bibnamefont {Rangelov}}, \bibinfo {author} {\bibfnamefont {B. W.}\
  \bibnamefont {Shore}}, \ and\ \bibinfo {author} {\bibfnamefont {K.}\
  \bibnamefont {Bergmann}},\ }\bibfield  {title} {\enquote {\bibinfo {title}
  {{Stimulated Raman adiabatic passage in physics, chemistry, and beyond}},}\
  }\href {\doibase 10.1103/RevModPhys.89.015006} {\bibfield  {journal}
  {\bibinfo  {journal} {Rev. Mod. Phys.}\ }\textbf {\bibinfo {volume} {89}},\
  \bibinfo {pages} {15006} (\bibinfo {year} {2017})}\BibitemShut {NoStop}%
\bibitem [{Note2()}]{Note2}%
  \BibitemOpen
  \bibinfo {note} {This error resulted from the wrong treatment of the
  branching ratios, $W_{\Omega \rightarrow X}$, from an initial, $\Omega $, to
  a final channel, $X$. This assumption lead to erroneous autoionization widths
  $\Gamma _\Omega (r)$ only dependent on the initial channel. In this work, the
  autoionization widths depend on both the initial and the final channel by
  means of the branching ratios, $\Gamma _{\Omega \rightarrow X}(r)=\Gamma
  _\Omega (r) W_{\Omega \rightarrow X}$}\BibitemShut {NoStop}%
\bibitem [{\citenamefont {Mori}\ \emph {et~al.}(1964)\citenamefont {Mori},
  \citenamefont {Watanabe},\ and\ \citenamefont {Katsuura}}]{Mori1964}%
  \BibitemOpen
  \bibfield  {author} {\bibinfo {author} {\bibfnamefont {M.}~\bibnamefont
  {Mori}}, \bibinfo {author} {\bibfnamefont {T.}~\bibnamefont {Watanabe}}, \
  and\ \bibinfo {author} {\bibfnamefont {K.}~\bibnamefont {Katsuura}},\
  }\bibfield  {title} {\enquote {\bibinfo {title} {{Collisional excitation
  transfer between atoms in the resonant process}},}\ }\href {\doibase
  10.1143/JPSJ.19.380} {\bibfield  {journal} {\bibinfo  {journal} {J. Phys.
  Soc. Japan}\ }\textbf {\bibinfo {volume} {19}},\ \bibinfo {pages} {380}
  (\bibinfo {year} {1964})}\BibitemShut {NoStop}%
\bibitem [{\citenamefont {Zare}(1988)}]{Zare1988}%
  \BibitemOpen
  \bibfield  {author} {\bibinfo {author} {\bibfnamefont {R.~N.}\ \bibnamefont
  {Zare}},\ }\href@noop {} {\emph {\bibinfo {title} {{Angular Momentum:
  Understanding Spatial Aspects in Chemistry and Physics}}}}\ (\bibinfo
  {publisher} {John Wiley and Sons},\ \bibinfo {address} {New York, USA},\
  \bibinfo {year} {1988})\BibitemShut {NoStop}%
\end{thebibliography}

%

\end{document}